\documentclass[preprint, prl, floatfix]{revtex4-1}
\usepackage{multirow}
\usepackage{amsmath}
\usepackage{color}
\usepackage{graphicx}  
\usepackage{tabularx, array}
\usepackage[normalem]{ulem} 
\graphicspath{{./FIGURES/}}

\begin{document}

\title{Diffusion in jammed particle packs}

\author{Dan S. Bolintineanu}
\email{dsbolin@sandia.gov}
\affiliation{Sandia National Laboratories, Albuquerque, New Mexico 87185}
\author{Gary S. Grest}
\affiliation{Sandia National Laboratories, Albuquerque, New Mexico 87185}

\author{Jeremy B. Lechman}
\affiliation{Sandia National Laboratories, Albuquerque, New Mexico 87185}

\author{Leonardo E. Silbert}
\affiliation{Department of Physics, Southern Illinois University Carbondale,
  Carbondale, IL, 62901}

\date{\today}

\begin{abstract}

Using random walk simulations we explore diffusive transport through monodisperse sphere packings over a range of packing fractions, $\phi$, in the vicinity of the jamming transition at $\phi_{c}$. Various diffusion properties are computed over several orders of magnitude in both time and packing pressure. Two well-separated regimes of normal, ``Fickian'' diffusion, where the mean squared displacement is linear in time, are observed. The first corresponds to diffusion inside individual spheres, while the latter is the long-time bulk diffusion. The intermediate anomalous diffusion regime and the long-time value of the diffusion coefficient are both shown to be controlled by particle contacts, which in turn depend on proximity to $\phi_{c}$. The time required to recover normal diffusion $t^*$ scales as $(\phi-\phi_c)^{-0.5}$ and the long-time diffusivity $D_\infty \sim (\phi - \phi_c)^{0.5}$, or $D_\infty \sim 1/t^*$. It is shown that the distribution of mean first passage times associated with the escape of random walkers between neighboring particles controls both $t^*$ and $D_\infty$ in the limit $\phi \rightarrow \phi_c$.

\end{abstract}


\maketitle

Diffusive transport through heterogeneous media has been a long-standing area of interest~\cite{torquato2002random,sahimi2003heterogeneous}, with wide-ranging applications in biology, geophysics and materials science. Heterogeneous media consisting of packed particles are of particular interest both as model systems and for industrial applications that involve powder processing. Analogous processes are electrical conduction in particle packs, relevant for energy storage applications (i.e. battery electrodes) e.g., \cite{Chen01012013}; or molecular diffusion in porous materials with narrow inter-pore connections, relevant for biological applications (i.e. intercellular transport)\cite{Schuss09102007}.  Furthermore, the relationship between inhomogeneous material structure and bulk transport properties remains poorly understood in the context of multi-scale effects. Locally, material crystallline structure or interfacial effects can dominate transport, while at larger length and time scales, topology and connectivity prevail. Experimental investigations that simultaneously probe these length scales are challenging. In this work, we use detailed simulations to explore the relationship between the bulk particle packing fraction $\phi$, which can be measured and controlled experimentally; microstructural features that govern transport, such as the interparticle contact area; and diffusive transport over a broad range of length and time scales.

A rich literature exists for effective medium theories of transport properties in heterogeneous media~\cite{landauer1952electrical, bruggeman1935berechnung, maxwell1881treatise, torquato1985effective, torquato2002random, sahimi2003heterogeneous}, while much is known about diffusion on regular and disordered lattices~\cite{hausandkehr1987} as well as anomalous diffusion in disordered media~\cite{bouchaudandgeorges,havlinandbenavraham} from a stochastic process point of view. In addition, work on percolation theory and in particular continuum percolation has been directed at transport properties of random structures~\cite{staufferandaharony1994,hughes2009,balberg2009}. More recently, simulations of ballistic and diffusive transport using the Lorentz model of random overlapping spheres have been performed~\cite{hoflingetal2006,hoflingetal2008}.  The results of that work are analyzed elegantly in a traditional continuum percolation-type framework (the so-called swiss cheese problem).  Here, we are interested in conduction through the particle matrix and more importantly in systems that deviate from standard percolation (i.e. fractal) structures but that still show interesting scalings in the long-time limit (i.e. hyperuniform structures)~\cite{donevetal2005}.  

To this end, we focus on compressed, disordered packings of monodisperese spheres over a range of packing fractions $\phi$, in the vicinity of the jamming transition~\cite{liu1998nonlinear} that takes place at a critical packing fraction $\phi_{c}$. $\phi$ represents a convenient macroscale parameter that controls the nature of interparticle contacts: below $\phi_{c}$, packings lose mechanical stability, whereas for $\phi > \phi_{c}$, a number of scaling relations appear that determine various structural and mechanical properties of the packing~\cite{PhysRevLett.84.4160,  ohern2003jamming, PhysRevE.73.041304}. Numerous studies~\cite{0953-8984-22-3-033101} have highlighted the strong heterogeneity in the force or contact network. Such aspects of jammed packings lead to a number of unusual, or anomalous, mechanical and transport properties ~\cite{PhysRevE.70.061302,PhysRevLett.102.038001}.

\begin{figure}
\centering
\includegraphics[width=0.23\textwidth]{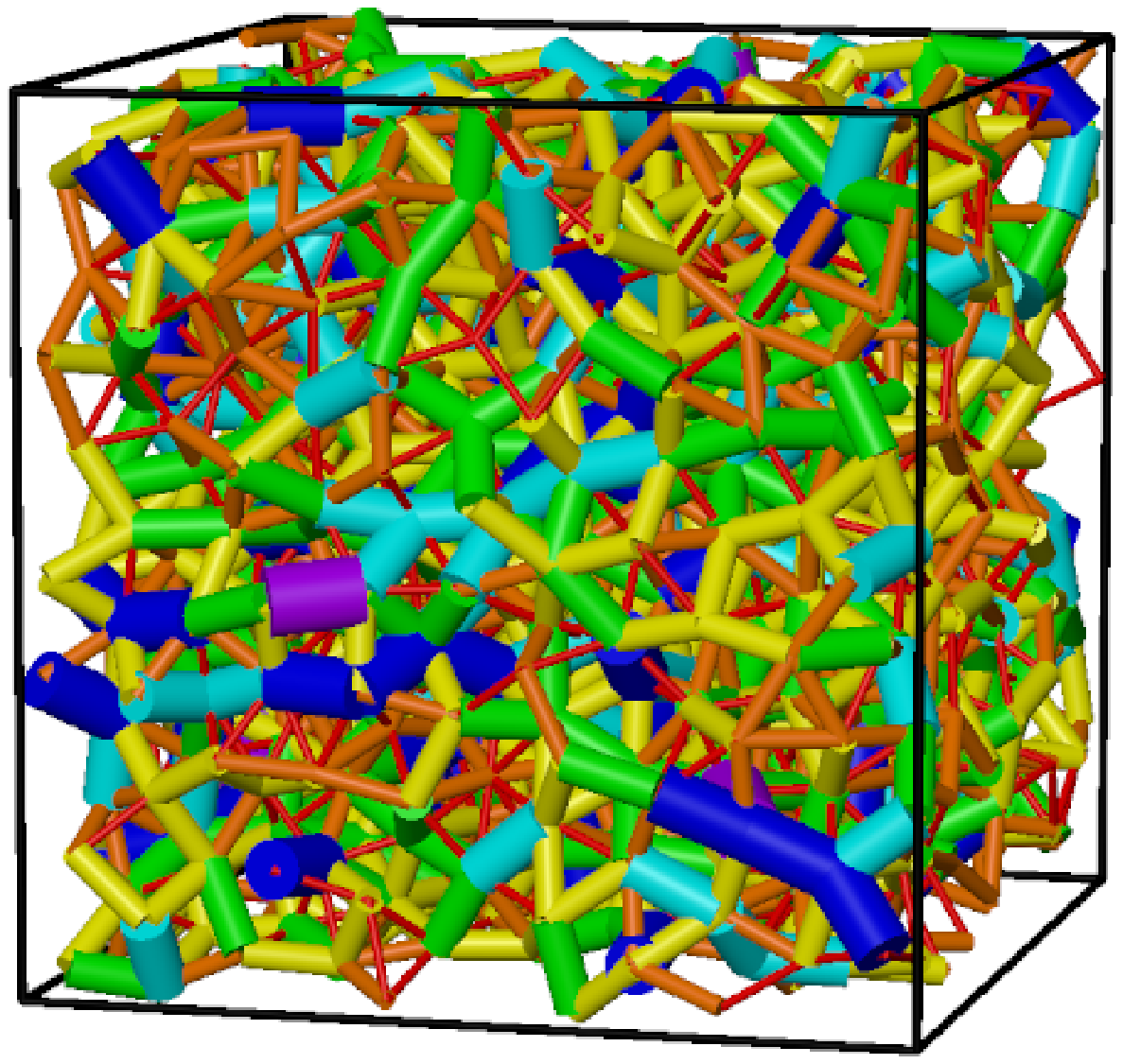}
\includegraphics[width=0.23\textwidth]{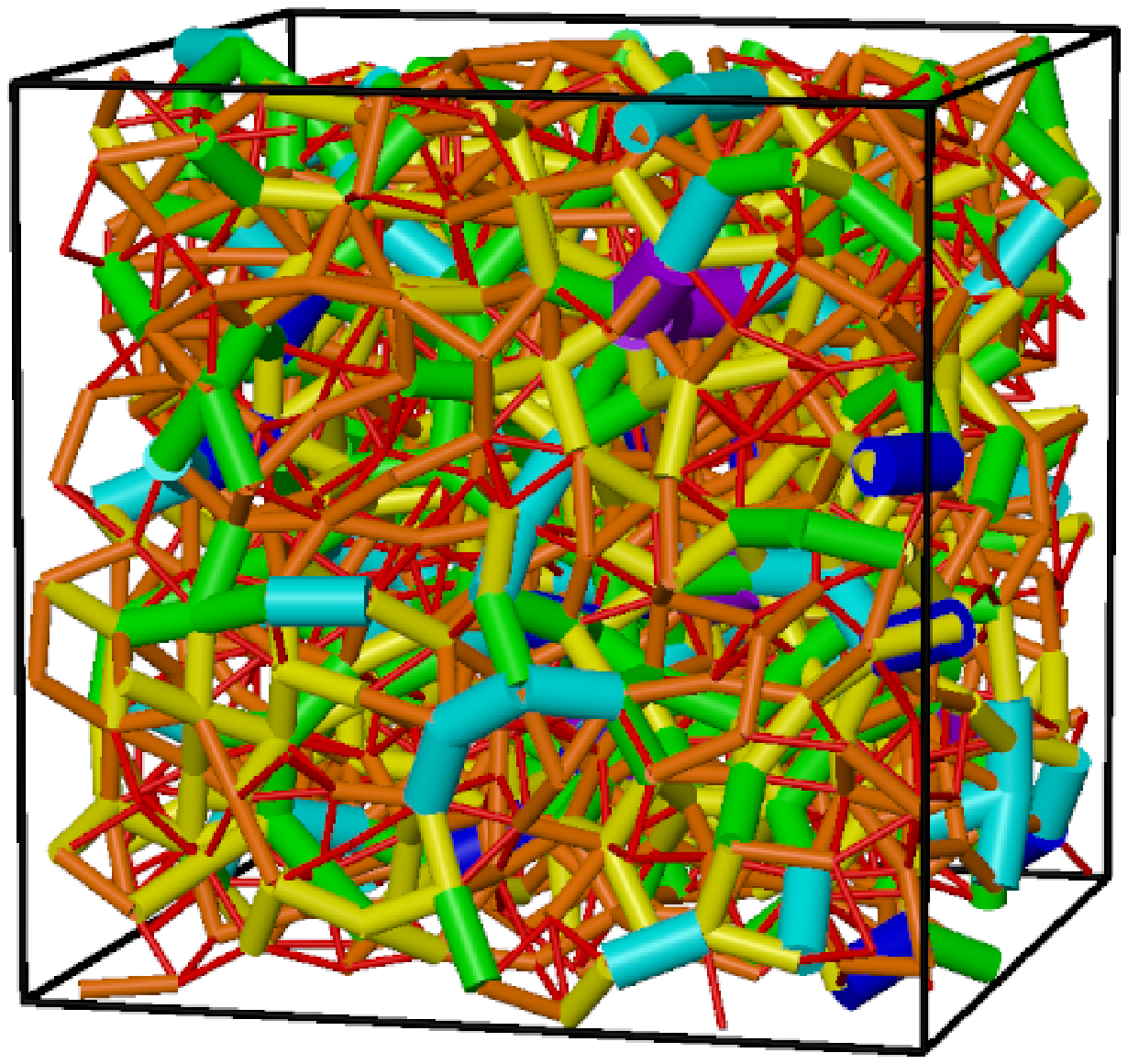}
\caption{Disordered, jammed packings of monodisperse spheres: far from   $\phi_c$ (left; packing pressure = $4.25 \times 10^{-2}$) and near $\phi_c$   (right; packing pressure = $4.29 \times 10^{-4}$). Lines represent contacts   between particles (not shown) with the thickness/shading indicating contact  area (relative to the largest contact for that system).} \label{fig:jammed_packs_network}
\end{figure}

Here, we take an explicit approach to exploring transport through jammed packings. We employ a stochastic technique based on a random walk algorithm~\cite{Kim_Torq_1989a} to track random walkers as they diffuse through the particle phase (equivalent to assuming a perfectly insulating surrounding medium), so that conduction cannot take place in the interstitial space. The interparticle contacts therefore play a crucial role in controlling the effective diffusivity, particularly as $\phi \rightarrow \phi_c$, where the interparticle contact areas are small compared to the particle size. This key feature is illustrated in Fig.~\ref{fig:jammed_packs_network}, which shows two packings prepared at different packing pressures, or equivalently $\Delta\phi \equiv \phi - \phi_{c}$, where lines of thickness corresponding to the size of the contact areas join the centers of particles. For compressed systems far from the jamming threshold, not only is the average coordination number larger, but the contacts themselves have larger overlap areas. Closer to the jamming transition, contact areas decrease, and one would therefore expect diffusion through the packings to depend on proximity to the jamming transition.

Indeed, we observe two distinct diffusion domains, one governed by diffusion of random walkers inside the particles, and the other by rare crossings of random walkers between neighboring particles. As $\phi \rightarrow \phi_c$, the separation in time scales for these domains diverges, and inter-particle hopping becomes the dominant, transport-limiting process. We therefore analyze the transport problem in the context of narrow escape/mean first passage time~\cite{Redner_book}. A recent analysis of diffusion from a sphere with multiple, well-separated, small absorbing windows~\cite{Cheviakov_MFPT} provides the essential relationships between first passage times and interparticle contact areas. Small changes in packing fraction, which affect interfacial contact areas between particles, ultimately lead to changes of several orders of magnitude in bulk properties. Variations in interfacial areas in the current case of spheres with perfect contact are analogous to variations in interfacial contact resistance (in such systems, these could be controlled by system pressure, sintering processes, or particle surface treatments). Characterizing this contact resistance and connecting it to controllable bulk properties in the way we connect interfacial area to packing pressure is highly system-dependent; however, despite the idealized nature of the present case, these phenomena apply to a broad range of important systems.

Our systems are composed of overcompressed, mechanically stable particle packings prepared using an established (de)compression protocol~\cite{silbert2010jamming}. Each packing consists of $N$ monodisperse, inelastic, frictionless spheres of diameter $d$ and mass $m$ in a cubic simulation box with periodic boundary conditions. All results presented here are in dimensionless units in which $m=1$ and $d=1$. Different packings were generated with particles interacting only on contact through a Hookean force law (spring force constant $k = 1$). By adjusting the packing fraction $\phi$ over small intervals, multiple packings were generated at precise values of the packing pressure $P$, over the range $4 \times 10^{-8} < P < 4 \times 10^{-1}$. For the systems studied here, we found $\phi_{c} \approx 0.6383$, which is consistent with the value of random close packing for frictionless spheres\footnote{In contrast, frictional particles tend to pack at lower densities all the way down to the random close packing value of approximately 0.55~\cite{silbert2010jamming}; for comparison, the packing fraction of typical desert sand is between 0.6-0.7}. Our major results are for $N=1000$, but we also checked that system size did not affect our results using $N=10000$, as shown in the Supplementary Material. One consequence of this packing generation protocol is the presence of rattlers - particles (or clusters of contacting particles) that have no contact neighbors. For typical jammed configurations close to $\phi_{c}$, less than 5\% of particles are rattlers, which were removed from the packing prior to random walk simulations.

\begin{figure}[h]
  \centering 
  \includegraphics[width=0.4\textwidth]{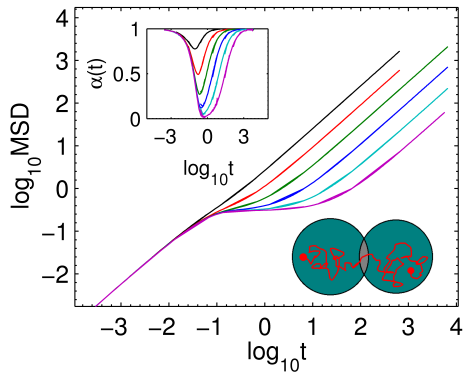}
  \caption{Mean squared displacement (MSD) of random walkers in jammed packings spanning $10^{-6} < \Delta\phi < 10^{-1}$. Insets: The anomalous diffusion exponent $\alpha (t)$ and a schematic of the random walker simulation. Color legend is the same as in Figure~\ref{fig:diff_jammed2}.}
  \label{fig:diff_jammed1}
\end{figure}

We employed a random walker algorithm similar to that developed by Kim and Torquato~\cite{Kim_Torq_1989a, Kim_Torq_1990, Kim_Torq_1991}. The basis of the algorithm is a continous-time, off-lattice random walk within a heterogeneous structure, illustrated in the inset to Fig.~\ref{fig:diff_jammed1}. A key difference in our algorithm is that at each integration step, the time elapsed must be the same for all walkers. This leads to a slightly more complex simulation algorithm, but greatly simplifies the subsequent analysis of time-dependent properties. Specifically, all random walkers must have the same net displacement $\Delta x_o$ (and therefore same elapsed time) at any given integration step \footnote{The value of the  net displacement $\Delta x_o = 0.02d$ was used throughout and chosen based on computational cost and does not affect the results.}. At the start of each simulation, ten times as many random walkers as spheres are randomly placed in the interior of particles. For a given walker at the start of a step, the distance to the edge of the particle in which it is currently located, denoted $\Delta x_p$, is first computed. If this distance is greater than $\Delta x_o$, the walker is translated a distance of $\Delta x_o$ in a randomly chosen direction, and the time step is complete. The time associated with this move is $\Delta x_o^2/6D_o$, where $D_o$ is the bulk diffusion coefficient inside the particles, which we set to unity with no loss of generality. However, if $\Delta x_i < \Delta x_p < \Delta x_o$, where $\Delta x_i$ is our minimal move resolution, the random walker is moved a distance $\Delta x_p$ in a randomly chosen direction, but the time step is not completed. The process is repeated until the walker's net displacement during the time step is exactly equal to $\Delta x_o$. This then completes the time step. If at some point $\Delta x_p < \Delta x_i$, the random walker is moved a distance of $\Delta x_i$ in a random direction; if this move results in the random walker being located outside of any particle, it is reflected back into the interior of the particle by a distance $\Delta x_i$ in a new random direction. The value of $\Delta x_i$ can affect the measured properties, and must be varied accordingly to resolve the smallest relevant length scale in the problem, which here is the size of the contact area (i.e. smaller $\Delta \phi$ require smaller $\Delta x_i$); tests of convergence with respect to $\Delta x_i$ are shown in the Supplementary Material.
 
\begin{figure}[h]
  \centering 
  \includegraphics[width=0.45\textwidth]{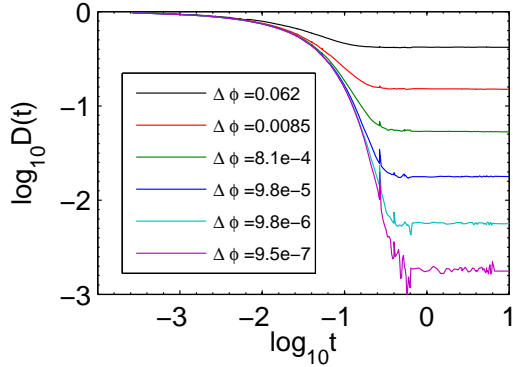}
  \caption{Evolution of the effective diffusion coefficient for the different
    packings.}
  \label{fig:diff_jammed2}
\end{figure}

Random walk simulations were carried out on the different packings for a wide range of pressures. Figures \ref{fig:diff_jammed1} and \ref{fig:diff_jammed2} show key measures characterizing the diffusion of random walkers in jammed particle packs for several values of $\Delta \phi$.

As seen from the plot of mean squared displacement (MSD) in Fig.~\ref{fig:diff_jammed1}, the short-time diffusion behavior is Fickian (MSD $\sim t$) for all $\Delta \phi$. This corresponds to diffusion of random walkers inside individual particles, on time scales shorter than that at which random walkers encounter the particle boundary. At intermediate times (starting at $t \sim 10^{-1}$), a subdiffusive plateau develops, which becomes broader as $\phi \rightarrow \phi_{c}$. At sufficiently long times, linear diffusive behavior is once again recovered for all systems as the walkers explore the entire packing. The time required to reach the long-time linear regime increases with decreasing $\Delta \phi$. This time-dependent diffusion behavior can be expressed either as a time-dependent effective diffusion coefficient $D(t)$, such that $\mathrm{MSD} = D(t)t/6$, or a time-dependent anomalous diffusion exponent $\alpha(t)$, $\mathrm{MSD}\sim t^{\alpha(t)}/6$. The inset of Fig.~\ref{fig:diff_jammed1} shows $\alpha (t) = d\log{\mathrm{MSD}}/d\log t$ for several values of $\Delta \phi$. The Fickian diffusive regimes at short and long times correspond to $\alpha \approx 1$. Alternatively, Fig.~\ref{fig:diff_jammed2} shows the time-dependent behavior of $D(t)$. At short times, $D(t) \sim D_o$, corresponding to the diffusion coefficient inside the particles. At longer times, $D(t)$ decreases due to the trapping of random walkers inside particles, which becomes more pronounced as $\phi \rightarrow \phi_c$ due to smaller interparticle contacts. The long-time plateau value $D_\infty$ corresponds to diffusion at time scales beyond those associated with interparticle hopping, and thus reflects the bulk homogeneous limit.

\begin{figure}
  \centering
  \includegraphics[width=0.22\textwidth]{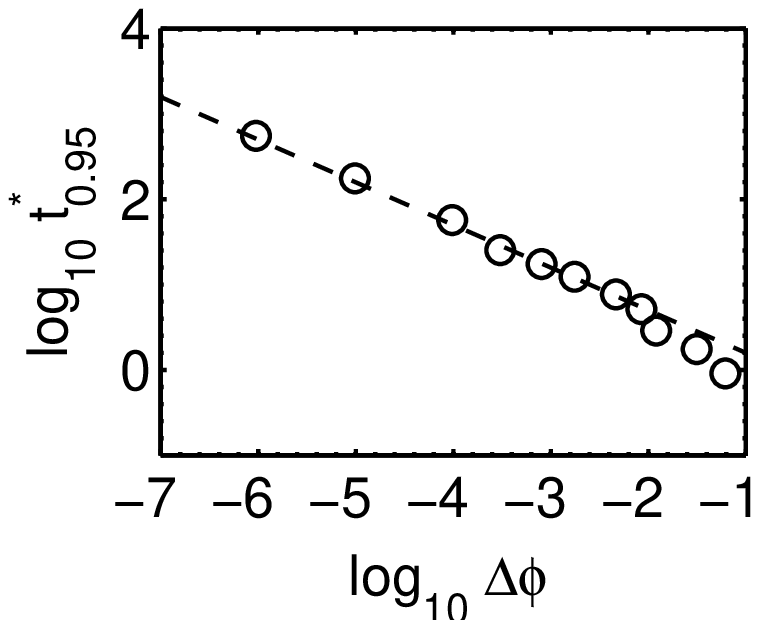}
  \includegraphics[width=0.22\textwidth]{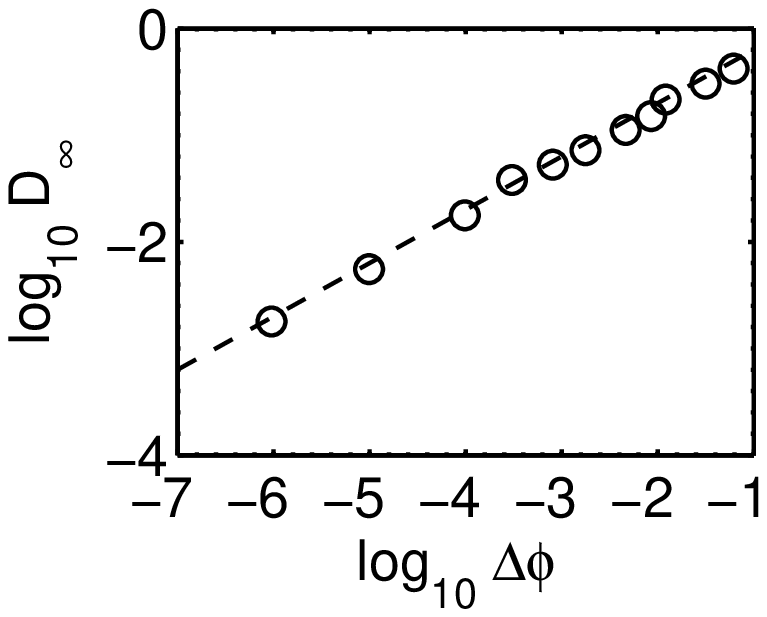}
  \caption{(a) Characteristic time $t_{0.95}^{*}$ at which Fickian diffusion is recovered. Dashed black line corresponds to $t^*_{0.95} = 0.5(\phi - \phi_c)^{-0.5}$. (b) Late-time value of the diffusion coefficient $D_\infty$  as a function of $\Delta \phi$. Line corresponds to $D_\infty = 2.0(\phi - \phi_c)^{0.5}$}
  \label{fig:p_tstar_D_jammed}
\end{figure}

In order to quantify the transition between the various diffusion regimes, we define the characteristic time $t_{0.95}^*$ such that $\alpha(t_{0.95}^*) = 0.95$ near the transition from the intermediate subdiffusive regime to the Fickian regime. This represents the characteristic time at which bulk Fickian diffusion is recovered. $t_{0.95}^*$ is plotted in Figure~\ref{fig:p_tstar_D_jammed}(a) as a function of $\Delta \phi$. A clear scaling of $t^* \sim (\phi-\phi_{c})^{-0.5}$ is observed, with perhaps the exception of large $\Delta \phi$ values, which lie outside of the traditional jamming-scaling regime~\cite{ohern2003jamming}. Similarly, in Fig~\ref{fig:p_tstar_D_jammed} (b), the long-time plateau value of the diffusion coefficient (late-time values in Fig~\ref{fig:diff_jammed2}) is shown as a function of $\Delta \phi$. Note a clear scaling of $D_\infty \sim (\phi - \phi_{c})^{0.5}$.

To understand the scaling behavior observed for $t^*$ and $D_\infty$, we analyze the diffusion of random walkers in the context of a narrow escape, or first passage time~\cite{Redner_book}. Here the first passage time is the time for a random walker to escape the interior of a given particle to any of its neighboring particles. To escape, a random walker must reach the small circular contact areas that separate a particle from its neighbors. The analogous problem has been studied by Cheviakov et al~\cite{Cheviakov_MFPT}, who derived an asymptotic analytical solution for the mean first passage time $\mathrm{MFPT}(\mathbf{x})$ as a function of the starting position $\mathbf{x}$ and the size and relative orientation of small circular openings on the surface of a spherical particle. The analytical solution requires that the radius of the circular openings $r_c$ be much smaller than the radius of the particle $r$ and that they are well separated~\cite{Cheviakov_MFPT}. These conditions are satisfied for all cases of interest here. For a given particle, we compute the volume-averaged MFPT, i.e. the MFPT averaged over all starting positions $\mathbf{x}$ and random walk realizations inside the particle (for a full description, see the work of Cheviakov et al~\cite{Cheviakov_MFPT}, in particular eq.~2.44 of that work). We refer to the volume-averaged MFPT as $\bar{t}$.

\begin{figure}
\centering
\includegraphics[width=0.22\textwidth]{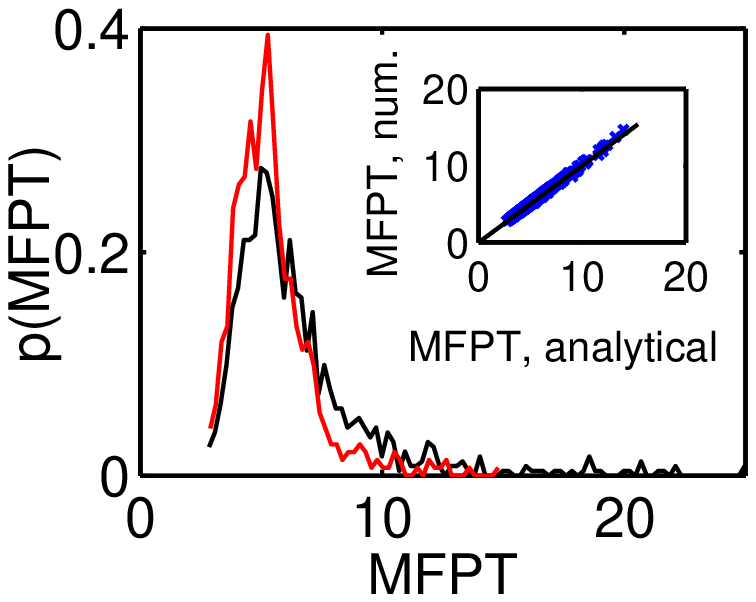}
\includegraphics[width=0.22\textwidth]{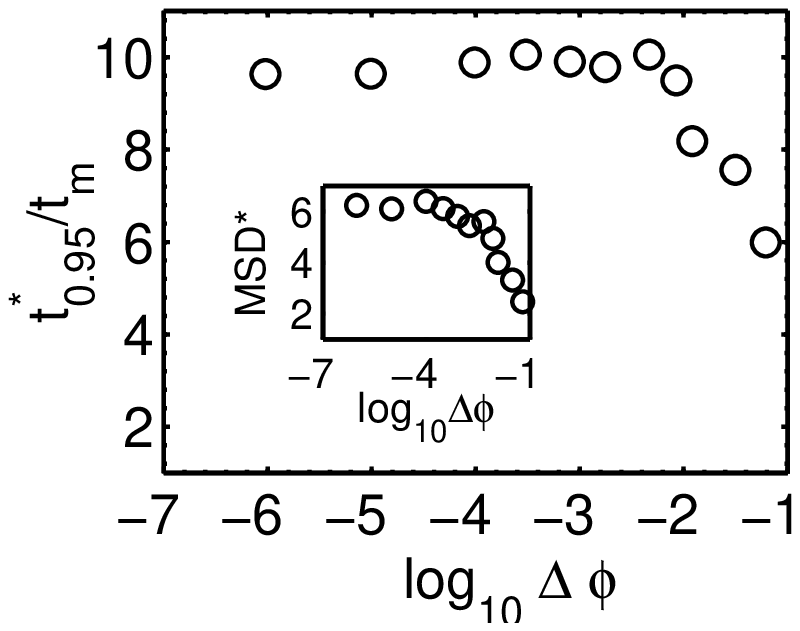}
\caption{(a) Comparison of volume-averaged MFPTs computed from simulations (red) and from the analytical expression\cite{Cheviakov_MFPT} (black) for $\Delta \phi = 9.8 \times 10^{-5}$.  The main figure shows the comparison of the histograms, the inset shows the direct comparison of values for all particles. The solid line in the inset has a slope of unity and zero intercept. (b) Relationship between $t^*_{95}/t_m$ and MFPT for all $\Delta \phi$. Inset shows corresponding $\mathrm{MSD}$ at $t=t^*_{95}$}
\label{fig:MFPT}
\end{figure}

Based on the structures of the packing configurations, we have computed $\bar{t}$ for $\Delta \phi=9.8 \times 10^{-5}$, and have compared this result to our random walk data in Fig.~\ref{fig:MFPT}(a). The agreement is very good; small discrepanices are due to convergence errors in the simulation data (for these simulations, convergence errors are much larger than other data previously discussed). Given this satisfactory correspondence, we use the analytical result of Cheviakov et al~\cite{Cheviakov_MFPT} in our subsequent analysis. For each system, the median value of $\bar{t}$, denoted $t_m$, is designated as the system average MFPT. In Figure~\ref{fig:MFPT}(b), we plot $t^*_{95}/t_m$ as a function of $\Delta \phi$. At all but the highest $\Delta \phi$ values, the ratio is constant, indicating that the characteristic time $t^*$, which signals the crossover from the subdiffusive to the long-time, bulk diffusive behavior is set by the MFPT, i.e. $t^* \sim t_m$. Hence, large-scale diffusion, as $\phi \rightarrow \phi_c$, is limited by the ability of a random walkers to escape from an average particle to its neighbor, which in turn is governed by the average size of the interparticle contacts.

The constant value of $\sim 10$ to which the $t^*_{95}/t_m$ ratio converges in Figure~\ref{fig:MFPT}(b) is not readily explicated, since it is convoluted by the fact that $t_m$ is a median value of the volume-averaged MFPT. Instead, we plot the MSD of the random walkers at $t=t^*_{95}$ in the inset of Fig~\ref{fig:MFPT}(b). This quantity, which we denote $\mathrm{MSD}^*$, converges to a constant value of $\sim$6 as $\Delta \phi$ decreases. This can be readily interpreted as the average distance, $\sqrt{\mathrm{MSD}^*}\sim 2.5$, that a random walker must travel before its trajectory exhibits regular diffusion characteristics. In other words, in the limit $\phi \rightarrow \phi_c$, where diffusion is dominated by rare event interparticle hopping, $\sim$2-3 such hops are needed to reach the diffusive regime. Thus, the anomalous diffusive behavior is localised on the particle scale and not the direct result of any large length scale phenomenon. It is interesting to note, however, that this is consistent with the distance at which the radial distribution function, $g(r)$, loses history of the local packing structure and is likely consistent with the idea that the directionality of the discrete contact angles between particles is averaged out.

We have established that $t^* \sim t_m$, and Fig.~\ref{fig:p_tstar_D_jammed} shows that $t^* \sim t_m \sim (\phi - \phi_c)^{-0.5}$. To leading order, the analytical result~\cite{Cheviakov_MFPT} yields $\bar{t} \sim r_c^{-1}$, and therefore $t_m \sim r_c^{-1}$, where $r_{c}$ is the radius of the contact region. By geometric arguments $(\phi - \phi_c) \sim \delta$ for small $\delta$, where $\delta$ is the overlap between contacting particles $i$ and $j$, $\delta = d - r_{ij}$, and $r_{ij}$ is their center-center separation. Similarly, $r_c \sim \delta^{0.5}$ for small $\delta$. Combining all of these, we find:

\begin{equation}
  t^* \sim t_m \sim r_c^{-1} \sim \delta^{-0.5} \sim (\phi - \phi_c)^{-0.5}, 
\label{eq2}
\end{equation}
which is consistent with the results of Fig.~\ref{fig:p_tstar_D_jammed}(a)
\footnote{It is worth pointing out that this scaling of $t^{*}$ is consistent with the scaling of the characteristic frequency of the so-called Boson peak found in Ref.~\onlinecite{PhysRevLett.95.098301}. However, it is unclear whether there is any direct connection between these timescales.}.

The scaling behavior of the long-time effective diffusion coefficient $D_\infty$ can also be readily explained by a similar argument. In the long-time limit, normal diffusion is recovered, such that $\mathrm{MSD}\sim t$. For any $t^*$ value selected so that $\alpha$ is arbitrarily close to unity, the corresponding $\mathrm{MSD}^*$ will be constant at low $\Delta \phi$ values (see inset of Figure~\ref{fig:MFPT}(b) for $t^*_{0.95}$). The long-time diffusion coefficient in this case is then simply $D_\infty = \mathrm{MSD}^*/6t^*$, and so $D_\infty \sim 1/t^{*}$. Since $t^* \sim (\phi - \phi_c)^{-0.5}$, it follows that $D_\infty \sim (\phi - \phi_c)^{0.5}$, the scaling behavior noted in Figure~\ref{fig:p_tstar_D_jammed}(b).

In summary, we have presented random walk simulations of diffusive transport in disordered particle packings in the vicinity of the jamming transition. Two distinct linear diffusive regimes where $\mathrm{MSD} \sim t$ are observed at short and long time scales, separated by an intermediate subdiffusive regime. The first regime corresponds to regular diffusion in the interior of particles, while the second regime is the expected long-time bulk diffusion behavior. The time scale separation between the two regimes grows as the volume fraction approaches the critical jamming value. This is a result of the corresponding decrease in the size of interparticle contacts, which leads to the average diffusion behavior being controlled by the ability of random walkers to escape from one particle to neighboring coordinated particles, so-called narrow escape. This behavior can be analyzed in the context of a mean first passage time. We observe scaling of $t^{*} \sim (\phi - \phi_c)^{-0.5}$ and $D_\infty \sim (\phi - \phi_c)^{0.5}$, where $t^{*}$ is a characteristic time at which linear diffusion is recovered and $D_\infty$ is the long-time effective bulk diffusion coefficient. Since nearest-neighbor inter-particle contacts control the large-scale transport properties in the particle packs considered here, this behavior is not unique to systems near the jamming point since it is not the presence of disorder that controls the anomalous diffusive behavior but rather the size of the particle contacts themselves. Therefore, we expect \cite{longpaper} the qualitative trends, as well as the scaling relations reported here, to be independent of the interparticle potential as well as packing geometry. It is hoped that this work will motivate detailed experimental studies of tranport dynamics at intermediate scales in systems dominated by interfacial resistance, e.g. pulse thermography~\cite{shepard2007automated, fournier1989heterogeneous} in particle packs.

\section*{Acknowledgements}
 
This work was supported by the Sandia Laboratory Directed Research and Development Program. This work was performed, in part, at the Center for Integrated Nanotechnologies, a U.S. Department of Energy, Office of Basic Energy Sciences user facility. Sandia National Laboratories is a multiprogram laboratory managed and operated by Sandia Corporation, a Lockheed-Martin Company, for the U. S. Department of Energy's National Nuclear Security Administration under Contract No. DE-AC04-94AL85000. L.E.S. gratefully acknowledges the support of PO 1355262 during a hospitable stay at SNL.
\bibliography{rwalk}

\pagebreak
\newpage
\clearpage

\end{document}